\documentclass[10pt,conference, compsocconf,fleqn]{IEEEtran}
\usepackage[utf8]{inputenc}
\usepackage[T1]{fontenc}
\usepackage{amsmath}
\usepackage{hyperref}
\usepackage{graphicx}
\usepackage{longtable}
\usepackage{wrapfig}
\usepackage{rotating}
\usepackage[normalem]{ulem}
\usepackage{amssymb}
\usepackage{capt-of}
\usepackage{listings}
\usepackage{xcolor}

\DeclareMathOperator*{\argmax}{arg\,max}
\DeclareMathOperator*{\argmin}{arg\,min}

\newcommand{\sut}{\mathcal{M}}
\newcommand{\R}{\mathbb{R}}

\usepackage{mathtools}
\usepackage[linesnumbered,ruled,vlined]{algorithm2e}

\author{
Jarkko Peltomäki and Ivan Porres\\
Faculty of Science and Engineering, \\ Åbo Akademi University \\ Turku, Finland \\
\texttt{name.surname@abo.fi}
}
\date{\today}
\title{Falsification of Multiple Requirements for Cyber-Physical Systems Using Online Generative Adversarial Networks and Multi-Armed Bandits}

\begin{document}

\maketitle
\begin{abstract}
  We consider the problem of falsifying safety requirements of Cyber-Physical Systems expressed in signal temporal
  logic (STL). This problem can be turned into an optimization problem via STL robustness functions. In this paper,
  our focus is in falsifying systems with multiple requirements. We propose to solve such conjunctive requirements
  using online generative adversarial networks (GANs) as test generators. Our main contribution is an algorithm which
  falsifies a conjunctive requirement $\varphi_1 \land \cdots \land \varphi_n$ by using a GAN for each requirement
  $\varphi_i$ separately. Using ideas from multi-armed bandit algorithms, our algorithm only trains a single GAN at
  every step, which saves resources. Our experiments indicate that, in addition to saving resources, this multi-armed
  bandit algorithm can falsify requirements with fewer number of executions on the system under test when compared to
  (i) an algorithm training a single GAN for the complete conjunctive requirement and  (ii) an algorithm  always
  training $n$ GANs at each step.
\end{abstract}

\section{Introduction}
\label{sec:intro}

In this paper, we study the problem of black-box falsification of  safety requirements of Cyber-Physical Systems (CPS).  This is a validation method to increase the confidence that a system works as expected before it is taken into production. This is especially important with safety requirements of CPS, where a fault can lead to severe damage or injuries.

We focus on Cyber-Physical Systems where inputs and outputs are given as real-valued signals. We define the safety requirements of such systems as properties expressed in signal temporal logic (STL)~\cite{maler2004monitoring}. An example of such requirement is
\begin{equation*}
  (\square_{[0,30]} \textup{RPM} < 3000) \rightarrow (\square_{[0,4]} \textup{SPEED} < 35).
\end{equation*}
Informally, this states that if the RPM signal is below 3000 during the first 30 units of time, then SPEED signal should be below 35 for the first 4 units of time. We test the system against these requirements by generating concrete input signals aiming to obtain an output signal that falsifies the given STL properties. By doing this, the input signals become a witness that the system under study does not fulfill the provided requirements.

Black-box falsification methods only consider inputs and observable outputs and do not require access to the internals of the system under study nor to its design specifications or source code. Black-box methods can be used at any point of the development process and only require that we are able to evaluate inputs and output pairs by simulation or actual execution. An example of a simulation environment is MATLAB. It is a popular design and simulation toolset for  Cyber-Physical Systems and it is widely used in academia and industry. Therefore many falsification tools, including the implementations of the algorithms presented here, can use it as a simulation environment.

The challenge in black-box safety validation of CPS lies in how to achieve the falsification of the proposed requirements with a limited testing budget. Corso et al. have recently published a survey of algorithms tackling this problem~\cite{DBLP:journals/corr/abs-2005-02979}. This survey presents four main strategies studied in the literature: optimization, path planning, reinforcement learning, and importance sampling. 

Optimization-based falsification uses requirement robustness functions~\cite{donze2010robust} to drive the search of falsifying inputs. Intuitively a robustness function indicates how close a given input is falsifying a requirement. A robustness function returns a value $\leq 0$ for inputs that falsify the requirements, and it should yield progressively lower values when its inputs approach a falsification region. Given this, minimizing a robustness function will provide us with inputs that falsify the requirements, if they exist.

Different authors have proposed the use of existing optimization meta-heuristics such as genetic algorithms or simulated annealing to drive the search for falsification inputs~\cite{DBLP:journals/corr/abs-2005-02979} and also new algorithms that combine model learning with global and local search~\cite{DBLP:conf/case/MathesenYPF19}. It is characteristic to these earlier algorithms that they have no knowledge of how the robustness functions are derived from requirements. More recently Zhang et al.~\cite{zhang2019multiarmed} and Mathesen et al.~\cite{mathesen2021efficient} have proposed falsification algorithms that use information about the structure of the requirements to drive the falsification search. The work by Zhang et al. accepts requirements of the form  $\square_I(\varphi_1 \land \varphi_2)$ or  $\square_I(\varphi_1 \lor \varphi_2)$ while the more recent work of Mathesen can deal with more general requirements of the form $\varphi_1 \land \cdots \land \varphi_n$ \cite{mathesen2021efficient}.

In this paper, we continue this line of research and propose an algorithm for the robustness-based falsification of CPS that uses the online GAN framework to generate falsifying inputs. We have used the online GAN framework in the past for the problem of performance testing~\cite{ogan}. As the first contribution of this article, we show here how the online GAN framework can also be applied for the falsification of safety requirements with a competitive performance.

The second  contribution presented in this article is an extension of our base algorithm that uses specific knowledge on how to falsify a conjunctive requirement $\varphi_1 \land \cdots \land \varphi_n$ efficiently. We first show how to solve this problem by training a GAN for each requirement
$\varphi_i$ separately as proposed in~\cite{mathesen2021efficient}. We offer experimental evidence that such an algorithm outperforms an algorithm training a single GAN for the complete conjunctive requirement. The drawback of such an approach is that it requires the simultaneous training of $n$ different GANs. This problem is also present in~\cite{mathesen2021efficient} where the creation of $n$ different Gaussian process models is required at each step. 
  
Inspired by the multi-armed bandit problem, we address this issue by proposing a third algorithm which, after a warm up period, trains a single GAN at a time which saves resources. Our experiments indicate that, in addition to saving resources, it is possible that this variant falsifies requirements with fewer number of executions on the system under test.
  
We proceed as follows. In \autoref{sec:problem}, we introduce the problem of falsification of requirements of CPS
formally and discuss related work in more detail. \autoref{sec:alg1} presents our initial algorithm for a single
conjunctive requirement falsification based on the online GAN framework. This algorithm is evaluated in
\autoref{sec:experiment1} using the Automatic Transmission Controller model, a benchmark problem presented
in~\cite{ernst2020archcomp}. We continue in \autoref{sec:mp} by extending our original algorithm to deal with
conjunctive requirements more efficiently and propose two new algorithms. The new algorithms are evaluated in
\autoref{sec:evaluation} using a synthetic problem proposed in~\cite{mathesen2021efficient}. Finally we present our
concluding remarks in the last section.

\section{Falsification of Formal Requirements of CPS}\label{sec:problem}

\subsection{Problem Description}

We assume that the safety requirements of the system under test (SUT) can be expressed in signal temporal logic (STL)
\cite{maler2004monitoring} as formulas $\varphi_1$, $\ldots$, $\varphi_n$.  Falsifying a formula $\varphi$ means exhibiting a test such that the behavior
(signal or trajectory) of the SUT while executing the test violates $\varphi$. In our case, falsifying the safety
requirements amounts to falsifying the conjunctive requirement $\varphi_1 \land \cdots \land \varphi_n$.

More precisely, we represent the SUT as a model $\sut$ which takes as its input a test $t$ and outputs a (possibly
vector-valued) signal $\sut(t)$ for which the truth value of $\varphi_1 \land \cdots \land \varphi_n$ can be evaluated.
In order to search for a falsifying test $t$, we turn the problem into an optimization problem via robustness
functions. As described, e.g., in \cite{donze2010robust}, an STL formula $\varphi$ can be effectively transformed into
a real-valued robustness function $\rho_\varphi$ such that $\varphi$ evaluates to true for a signal $s$ if and only if
$\rho_\varphi(s) \geq 0$. Moreover, the robustness function has the following stability property: small changes to a
signal $s$ with robustness $\rho_\varphi(s)$ of high absolute value do not affect the truth value of $\varphi$ whereas
small changes when $|\rho_\varphi(s)|$ is small could change it. This property turns the problem of falsifying
$\varphi$ to that of minimizing $\rho_\varphi$. In other words, our task is to solve the optimization problem
\begin{equation}\label{eq:sp}
  \argmin_t \rho_\psi(\sut(t))
\end{equation}
where $\psi = \varphi_1 \land \cdots \land \varphi_n$. The elementary properties of robustness functions allows us to
write \eqref{eq:sp} as
\begin{equation}\label{eq:op}
  \argmin_t \min_{i = 1, \ldots, n} \rho_{\varphi_i}(\sut(t)).
\end{equation}
We remark that the success of this plan depends on how complicated the formulas $\varphi_i$ are. It is reasonable to
assume, for example, that the function predicates appearing in the formulas refer to locally Lipschitz continuous
functions. Observe also that continuous signals need to be discretized appropriately.

As pointed out in \cite{mathesen2021efficient}, it is often the case that evaluating $\sut(t)$ is slow, but computing
$\rho_\varphi(s)$ is fast. Therefore it makes sense to assume that we have all values $\rho_{\varphi_i}(\sut(t))$,
$i = 1, \ldots, n$, available as soon as a test $t$ has been executed on the SUT. With this assumption, we gain
knowledge as we can use all values $\rho_{\varphi_i}(\sut(t))$, $i = 1, \ldots, n$, instead of just observing their
minimum. The assumption also helps with the scale problem \cite{mathesen2021efficient,zhang2019multiarmed}. Indeed, the
value of $\rho_{\varphi_i}$ could have wildly different scale than $\rho_{\varphi_j}$, so $\rho_{\varphi_i}$ could
effectively mask any information in $\rho_{\varphi_j}$ if we only get to observe the minimum. It is worth remarking
that this problem can persist even after appropriate scaling.

\subsection{Previous Work}
Solving \eqref{eq:op} can be approached in different ways. Common optimization methods, like the cross-entropy method
\cite{sankaranarayanan2012falsification}, have been used; see the introduction of \cite{mathesen2021efficient} for more
references to prior works.

We are interested in the recent paper \cite{mathesen2021efficient} of Mathesen et al. where Bayesian optimization is
used. The main points of their minBO algorithm are as follows. Let us write $\rho_i(t)$ for $\rho_{\varphi_i}(\sut(t))$
for brevity. First sample tests randomly and execute them on the SUT to obtain a training data
\begin{equation*}
  (t_j, \rho_i(t_j), \ldots, \rho_n(t_j)), \quad j = 1, \ldots N,
\end{equation*}
and best test $t^*$ with
\begin{equation*}
  t^* = \argmin_{t_j, j=1,\ldots,N} \min_{i=1,\ldots,n} \rho_i(t_j).
\end{equation*}
Then a Gaussian Process \cite{gaussian_processes} is fitted for each $\rho_i$ using the above training data. This
yields an approximate model for each $\rho_i$. These models are used to figure out a test $t_{N+1}$ which is likely to
give smaller robustness values than $t^*$. The selection of $t_{N+1}$ is done by maximizing expected improvement as is
common in Bayesian optimization \cite{bayesian_optimization}. More precisely: a candidate test $t'_i$ with highest
expected improvement $\textup{EI}_i$ (with respect to $t^*$) is selected separately for each $\rho_i$, and the final
candidate $t_{N+1}$ is chosen to be $t'_k$ with $k = \argmax_{i=1,\ldots,n} \textup{EI}_i$. The test $t_{N+1}$ is
executed on the SUT and the results are added to the training data. The best test $t^*$ is updated if needed, and the
above is repeated until the execution budget is exhausted.

In order to evaluate the minBO algorithm described above, two experiments are proposed in \cite{mathesen2021efficient}.
The first experiment is artificial and concerns properties of predefined but complicated and nonlinear functions. The
second experiment concerns two industry benchmark models: the automatic transmission (AT) model of
\cite{ernst2020archcomp} and the ground collision avoidance system (GCAS) autopilot model for the F-16 fighter jet
\cite{heidlauf2018verification}. The findings of \cite{mathesen2021efficient} are that the minBO algorithm performs
always at least as well as a similar Bayesian optimization approach which has only access to the minimum $\min_i
\rho_i(t)$. The minBO algorithm performs statistically significantly better in cases where the scale problem is present
(both use cases in the first experiment and GCAS in the second experiment). The minBO algorithm is proposed as a
performant solution to the scale problem.

\section{The Online GAN Algorithm for Robustness-Based Falsification}~\label{sec:alg1}

In this section, we present an algorithm which solves \eqref{eq:sp} using online GANs. We falsify a single requirement
meaning we assume that we can only observe the minimum robustness of multiple requirements.

\subsection{Online GAN Training}\label{ssec:ogan_training}
At the heart of the algorithm is the idea of using a GAN for optimization. The idea is the same as in \cite{ogan} where
an online GAN is used for maximization. The idea is the following. Let $\varphi$ be an STL formula, and suppose that
the test robustness function $t \mapsto \rho_\varphi(\sut(t))$ takes values in $[0, 1]$. We deem $\varphi$ to be
falsified if there exists a test $t$ such that $\rho_\varphi(\sut(t)) = 0$.

In addition to the SUT, we have two components: a generator $G$ and a discriminator $D$. Both $G$ and $D$ are machine
learning models, and in this paper they are neural networks. The generator $G$ is a mapping from a latent space to the
space of tests, and the aim is to train $G$ in such a way that when the latent space is sampled uniformly, we obtain,
via the map $G$, tests $t$ for which the robustness $\rho_\varphi(\sut(t))$ is low. The discriminator $D$ simply learns
the map $t \mapsto \rho_\varphi(\sut(t))$.

Initially a random search is performed to obtain training data for $D$. Then we find new tests and train $G$ as
follows. We generate candidate tests with $G$ and estimate their robustness with $D$ (this avoids executing tests on
the SUT). Whenever a test with high estimated robustness is found, we execute it on the SUT, add the test and its
robustness to the training data, and train $D$ using this updated training data. We then sample randomly points $x_1$,
$\ldots$, $x_N$ from the latent space and form the artificial training data
\begin{equation*}
  (x_i, 0), \quad i = 1, \ldots, N.
\end{equation*}
Using this training data, we train the composite model $D \circ G$ with the model parameters of $D$ frozen. Since $0$
is the smallest value the robustness function can attain, this encourages the parameters of $G$ to change in such a way
that it generates tests which $D$ estimates to have low robustness. As more and more data is collected, $D$ should
become more accurate and $G$ should thus generate tests with low robustness. Ideally $G$ generates a test with
robustness $0$ provided that $\varphi$ is falsifiable.

Notice that our approach does not follow the traditional GAN training \cite{goodfellow2014generative} where the
discriminator is trained to distinguish between fake and real samples. The traditional approach is not possible here as
we need to find the tests online.

Our online GAN approach can be seen as an instance of the idea of studying a SUT via a surrogate model which is
refined over time as more information is available. The minBO algorithm of \cite{mathesen2021efficient} fits into the
same framework: their surrogate model is a Gaussian process instead of an online GAN.

\subsection{Online-GAN Algorithm for a Single Conjunctive Requirement}
Here we present an online GAN algorithm which attempts to falsify a single STL formula $\varphi$ with robustness
function $\rho$ defined by $\rho(t) = \rho_\varphi(\sut(t))$. In our context of falsifying the safety requirements of a
SUT, the formula can be thought to be $\varphi_1 \land \cdots \land \varphi_n$, that is, we can only observe the minimum
robustness. The algorithm is presented in detail in \autoref{alg1}. We remark that finding the robustness function
$\rho$ is straightforward and implementations can be found in \cite{nickovic2020rtamt,cralley2020tltk}.

\begin{algorithm}
\SetAlgoLined
T := Latin hypercube sampling(initial sample size)\;
GAN := new model with generator GN and discriminator DN\;
\Repeat{\textup{outcome} $\leq 0 \lor |\textup{T}| = \textup{budget}$}{
   train(GAN, T)\;
   target := $0$\; 
   \Repeat{\textup{predict}(\textup{DN}, \textup{t}) $\leq$ \textup{target}}{
   target := target  + $\Delta$\; 
   t := generate(GN)\; 
    }
   outcome := \textup{robustness}(t)\;
   T := T $\cup$ \{(t, outcome)\}\; 
   }
\textbf{result} test suite T
 \caption{Online GAN test generation algorithm for a single requirement.}\label{alg1}
\end{algorithm}

The algorithm simply does what is described on a higher level in \autoref{ssec:ogan_training}. The first task is to perform a random search using a small part of our testing budget. In this algorithm we propose to use Latin hypercube sampling~\cite{DBLP:journals/ress/HeltonD03} for this purpose. After that, the algorithm proceeds with the search driven by the GAN. The generator is used to
find a test with high estimated fitness in the inner loop. Initially we set the target estimated fitness to be $0$
(falsified) but, as it is not necessarily possible to generate such a test (especially just after the algorithm has
started), we raise the target on each execution of the loop until a candidate test is found. As more training data is
available, we should be able to achieve lower and lower targets.

\section{Experiment 1: Automatic Transmission (AT) Model}\label{sec:experiment1}
In this section, we evaluate \autoref{alg1} on the automatic transmission (AT) model which is standard benchmark model
proposed in \cite{hoxha2015benchmarks}. The model performs automatic gear selection for a car when two input signals
throttle and brake are provided to it. The model outputs two signals: engine speed (in RPM) and vehicle speed (in mph).
\begin{figure}
  \includegraphics[width=\columnwidth]{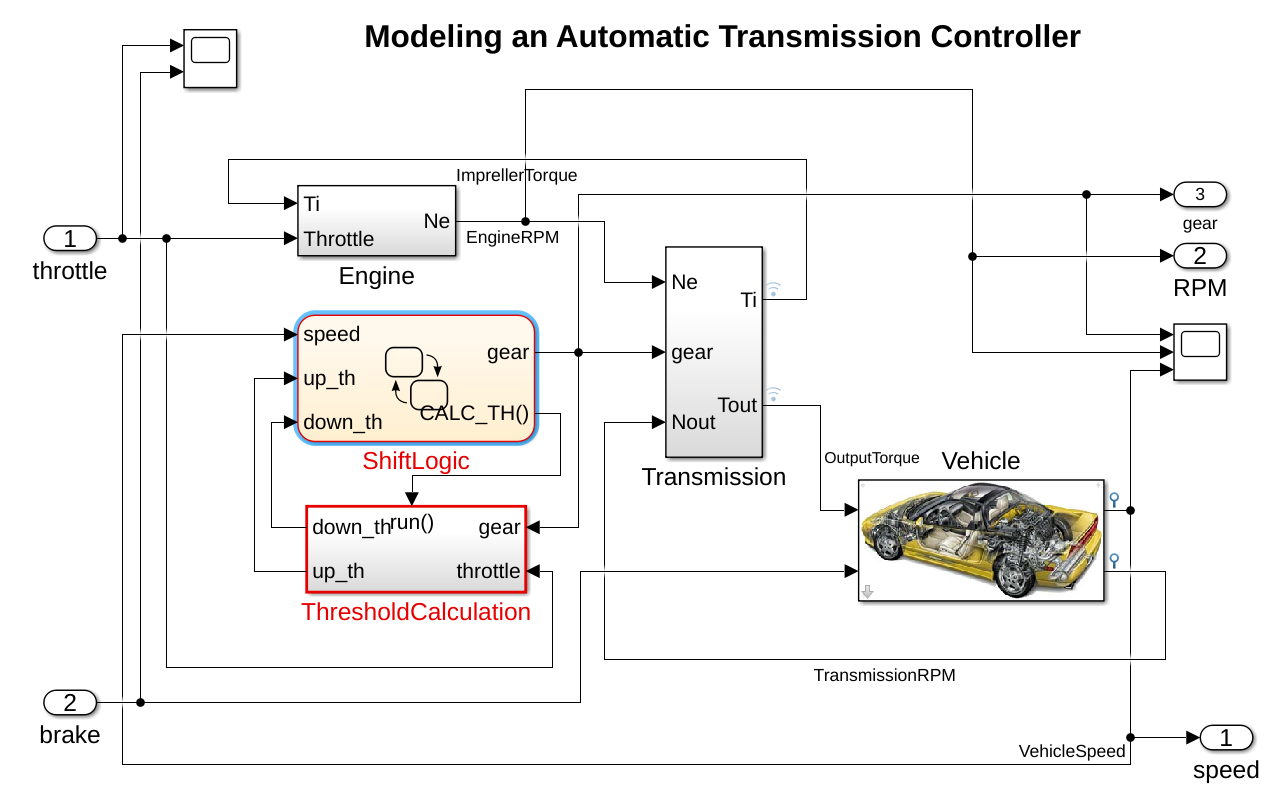}
  \caption{Overview of the MATLAB model for the Automatic Transmission Controller (AT) problem. Model source: \protect\cite{ernst2020archcomp}.}\label{fig:at-model}
\end{figure}

We are interested falsifying the requirement $\varphi_1 \land \varphi_2 \land \varphi_3$ where
\begin{align*}
  \varphi_1 &= (\square_{[0,30]} \textup{RPM} < 3000) \rightarrow (\square_{[0,4]} \textup{SPEED} < 35), \\
  \varphi_2 &= (\square_{[0,30]} \textup{RPM} < 3000) \rightarrow (\square_{[0,8]} \textup{SPEED} < 50), \\
  \varphi_3 &= (\square_{[0,30]} \textup{RPM} < 3000) \rightarrow (\square_{[0,20]} \textup{SPEED} < 65).
\end{align*}
These requirements are given, e.g., in the ARCH workshop 2021 competition \cite{ernst2021archcomp}. They require that
during the first $30$ time units (which is the complete duration of the signals) the initial vehicle speeds should take
values $35$, $50$, and $65$ provided that the engine speed is below $3000 \, \textup{RPM}$ during the whole execution.
The same falsification problem is considered in \cite{mathesen2021efficient}.

As in \cite{ernst2021archcomp}, we assume that $0 \leq \textup{THROTTLE} \leq 100$ and $0 \leq \textup{BRAKE} \leq 325$
during the whole execution (both signals can be positive simultaneously). Our input signals are piecewise constant
functions with $6$ pieces meaning that each input is constant for $5$ time units at a time. We discretize the signals
by sampling every $0.2$ time units. We represent our tests as vectors in $\R^{12}$ whose components satisfy the
preceding requirements.

A general way to find a robustness function for STL formulas described in \cite{donze2010robust}. However such a direct
approach is problematic as $\textup{RPM}$ and $\text{SPEED}$ are measured on very different scales. We use the
following ad hoc robustness function $\rho_1$ for $\varphi_1$:
\begin{equation*}
  \rho_1(\textup{RPM}, \textup{SPEED}) = \begin{cases}
                \tfrac12 (35 - M_{\textup{SPEED}})/35, & \!\!\!\! \text{if $M_{\textup{RPM}} < 3000$} \\
                M_{\textup{RPM}}/3000 - \tfrac12,      & \!\!\!\! \text{otherwise}
              \end{cases},
\end{equation*}
where
\begin{align*}
  M_{\textup{RPM}} &= \sup_{t \in [0,30]} \textup{RPM}(t) \text{ and} \\
  M_{\textup{SPEED}} &= \sup_{t \in [0,4]} \textup{SPEED}(t).
\end{align*}
We define robustness functions $\rho_2$ and $\rho_3$ analogously for $\varphi_2$ and $\varphi_3$. It is readily checked
that $\varphi_i$ is falsified if and only if there exists a signal $(\textup{RPM}, \textup{SPEED})$ such that
$\rho_i(\textup{RPM}, \textup{SPEED}) < 0$.

We attempt to falsify the requirement $\varphi_1 \land \varphi_2 \land \varphi_3$ using \autoref{alg1}. Due to the
stochastic nature of the algorithm, we repeat the falsification task $50$ times. We allow $80$ executions on the SUT
and use $25 \%$ of the execution budget for random search (Latin hypercube sampling). We have implemented
\autoref{alg1} in Python using TensorFlow for GAN implementation. The SUT is implemented in MATLAB as a Simulink model
available at the ARCH verification competition repository.\footnote{\url{https://gitlab.com/goranf/ARCH-COMP}} The
model is called via the MATLAB Python engine.

\begin{figure}
  \includegraphics[width=\columnwidth]{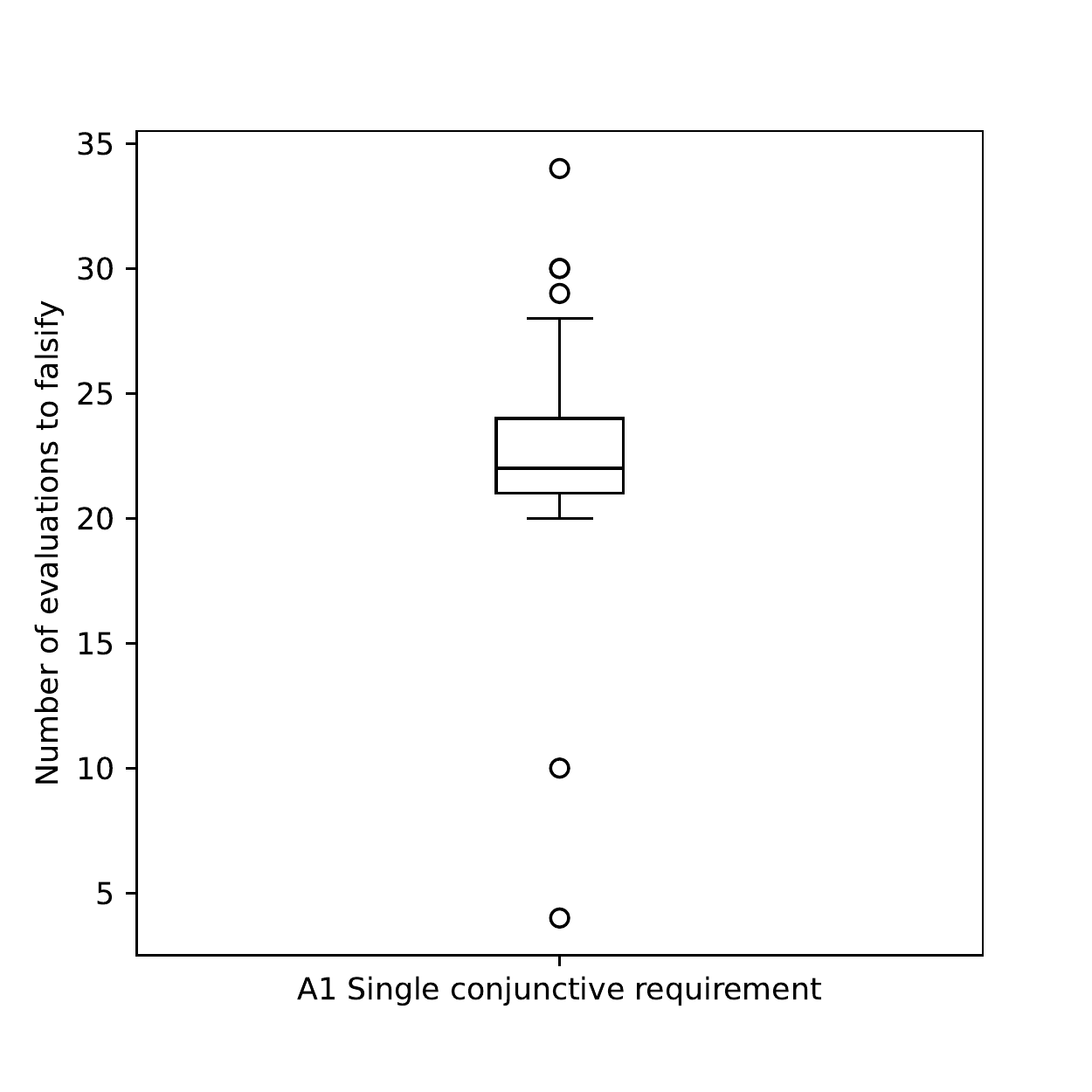}
  \caption{\textmd{Experiment 1: Box plot for number of executions needed for falsification.}}\label{fig:at-ffat}
\end{figure}

The box plot for the number of executions on the SUT required for falsification is found in \autoref{fig:at-ffat}.
\autoref{alg1} succeeded in falsification in each replication with mean $22.8$ (SD $4.4$) executions required for a
successful falsification. Since $20$ executions were allowed for random search, we conclude that typically the random
search was unable to falsify the requirement but the GAN could with just a few extra executions. This shows that
\autoref{alg1} is capable for falsification as intended.

Interestingly, the minBO algorithm of \cite{mathesen2021efficient} needed on average $68.7$ executions for
falsification. We explain the difference as follows. During research, we observed that the AT requirements are easily
falsified for constant input signals. When the GAN is trained for the first time, it is typical that the components of
its output are close to the middles of the allowed ranges. The first test proposed by the GAN is thus approximately
constant and has high chance succeeding in falsification. We take this as an indication that the tests proposed by the
minBO algorithm are not approximately constant.

We should note that we have not reproduced the results presented in~\cite{mathesen2021efficient} ourselves. Instead, we used the figures of the article, and therefore we cannot be certain that all the algorithms are evaluated under the same conditions.

\section{Online-GAN Algorithms for Multiple Requirements}
\label{sec:mp}
In this section, we present two Algorithms \ref{alg2} and \ref{alg3} which better take into account the information
on multiple requirements.

\subsection{Multiple Requirements}
In \autoref{alg2}, we present an online GAN algorithm which attempts to falsify multiple requirements $\varphi_1$,
$\ldots$, $\varphi_n$ with corresponding robustness functions $\rho_1$, $\ldots$, $\rho_n$.

\begin{algorithm}
\SetAlgoLined
T := Latin hypercube sampling(initial sample size)\;
GAN := nproperties new models with generators GN and discriminators DN\;
\Repeat{\textup{best} $\leq 0 \lor |\textup{T}| = \textup{budget}$}{
  \For{i $\in [1..\textup{nproperties}]$}{
   train(GAN, i, T)\;
   }
   target := $0$\; 
   \Repeat{\textup{p[best]} $\leq$ \textup{target}}{
     target := target  + $\Delta$; \\
     \For{\textup{i} $\in [1..\textup{nproperties}]$}{
       t[i] := generate(GN[i])\; 
       p[i] := predict(DN[i], t[i])\;
     }
     best := argmin(p)\;
   }
   outcome := robustness(t[best])\;
   best = min(outcome)\;
   T := T $\cup$ \{(t, outcome)\}\;
   }
\textbf{result} test suite T
 \caption{Online GAN test generation algorithm for multiple requirements.}\label{alg2}
\end{algorithm}

The difference to \autoref{alg1} is that we train $n$ online GANs, one for each requirement $\varphi_i$. When we search
for a candidate test, we consult the generators of each GAN and select the test which the corresponding discriminator
estimates to have the lowest robustness. Again, we raise the target robustness until a candidate test is found.

\subsection{Multiple Requirements with Property Selection}
Next we introduce a new idea which addresses an obvious problem with \autoref{alg2}: the requirements $\varphi_1$,
$\ldots$, $\varphi_n$ are not equal, so they should not be given the same amount of consideration. Indeed, if there is
a falsifiable requirement, then there is a requirement that is easiest to falsify and we should focus only on it. This
saves both generation time and executions on the SUT. The problem is, of course, that we do not know which requirement
(if any) is the easiest. Moreover, we are studying the requirements indirectly via the online GANs $\mathcal{G}_1$,
$\ldots$, $\mathcal{G}_n$ which act as surrogate models for the requirements and change over time. Our problem is thus
similar to the nonstationary multi-armed bandit (MAB) problem \cite{slivkins2019introduction}: how to explore all the
surrogate models $\mathcal{G}_i$ and how to exploit the ones we deem to be the best?

We propose the following simple approach for solving the problem. Whenever a candidate test is executed on the SUT, we
record which surrogate model $\mathcal{G}_i$ achieved the lowest robustness. Using these records we keep track of
success frequencies $p_1$, $\ldots$, $p_n$ for each surrogate model. When a new candidate test needs to be generated,
we proceed as follows: we pick a random surrogate model $\mathcal{G}$ according to the frequencies $p_1$, $\ldots$,
$p_n$ and use $\mathcal{G}$ to generate a new candidate test. In an initial warm-up period, we consult all surrogate
models in order to obtain good initial estimates for the success frequencies.

The described strategy clearly satisfies the requirements of exploration and exploitation: the historically most
successful surrogate model is most likely being selected again, but there is a chance that another is selected, which
perhaps leads to favoring an easily falsifiable requirement which initially looks unpromising. The success of this
approach obviously depends on the nature of the requirements $\varphi_1$, $\ldots$, $\varphi_n$ and the quality of the
initial random search. We leave it as an open problem to develop a better strategy which would take into account the
available information better. Recall that the minBO algorithm of \cite{mathesen2021efficient} uses expected improvement
to detect tests which likely have low robustness.

The variant of \autoref{alg2} with the above MAB-inspired approach is described in detail in \autoref{alg3}. On line
$3$, we pick a single GAN to be used for candidate generation based on the success frequencies. On line $13$, we record
which GAN achieved the lowest robustness when the selected test was executed on the SUT. Notice that this GAN might be
different from the GAN used for candidate test generation.

\begin{algorithm}
\SetAlgoLined
Apply \autoref{alg2} for N\% of the total budget. \\
\Repeat{\textup{best} $\leq 0 \lor |\textup{T}| = \textup{budget}$}{
   chosen := pick one from $[1..\textup{nproperties}]$ with weights winner$[1..\textup{nproperties}]$ \;
   train(GAN, chosen, T)\;
   target := $0$\; 
   \Repeat{\textup{p} $\leq$ \textup{target}}{
     target := target  + $\Delta$; \\
     t := generate(GN)\; 
     p := predict(DN, t)\;
   }
   outcome := robustness(execute(t))\;
   best, best\_index = min(outcome), argmin(outcome)\;
   winner[best\_index] := winner[best\_index] + 1\;
   T := T $\cup$ \{(t, outcome)\}\;
  }
\textbf{result} test suite T
 \caption{Online GAN test generation algorithm for multiple requirements with property selection.}\label{alg3}
\end{algorithm}

\section{Experiment 2: MO3d}\label{sec:evaluation}
In this section, we evaluate the three algorithms on a synthetic problem proposed in \cite{mathesen2021efficient}. It
concerns the nonlinear function $\textup{mo3d}\colon \R^3 \to \R^3, \textup{mo3d}(x) = (h_1(x), h_2(x), h_3(x))$ where
\begin{align*}
  h_1(x_1, x_2, x_3) &= 305 - 100 \sum_{i=1}^3 \sin\left(\frac{x_i}{3}\right), \\
  h_2(x_1, x_2, x_3) &= 230 - 75 \sum_{i=1}^3 \cos\left(\frac{x_i}{2.5}+15\right), \, \text{and} \\
  h_3(x_1, x_2, x_3) &= \sum_{i=1}^3 (x_i-7)^2 - \sum_{i=1}^3 \cos\left(\frac{x_i-7}{2.75}\right).
\end{align*}
This function achieves its componentwise minimum value at $x^* = (7,7,7)$ with $f(x^*) = h_3(x^*) = -3$. The other two
components achieve their minimum value of $5$ at the points $3\pi/2(1,1,1)$ and $-37.5(1,1,1)$ respectively.

We are interested in requiring that all elements of $f(x)$ should be greater than $0$ in the input domain $[-15,15]^3$.
This can be represented in STL as
\begin{equation*}
  (\square h_1(x) > 0) \land (\square h_2(x) > 0) \land (\square h_3(x) > 0).
\end{equation*}
We simply use the functions $h_1$, $h_2$, and $h_3$ themselves as robustness functions. We repeat the falsification
task $50$ times for each algorithm. We allow $80$ executions on the SUT and use $25 \%$ of the execution budget for
random search.

\begin{table}
  \caption{\textmd{Main Statistics of Experiment 2.}}\label{tbl}
  \begin{tabular}{c|c|c|c}
                                        & \textbf{A1}  & \textbf{A2} & \textbf{A3} \\ \hline
    \textbf{Falsifications after 50 repetitions}        &  $16$         & $39$        & $46$          \\
    \textbf{\% of falsifications}          &  $32 \%$          & $78 \%$        & $92 \%$           \\
    \textbf{Median observed minimum}         &  $2.84$          & $-1.36$        & $-2.01$           \\
    \textbf{Mean observed minimum }         &  $3.39$          & $-1.00$        & $-1.67$           \\
    \textbf{SD observed minimum}         &  $4.59$          & $1.72$        & $1.25$           \\
   \end{tabular}
\end{table}

\begin{figure}
  \includegraphics[width=\columnwidth]{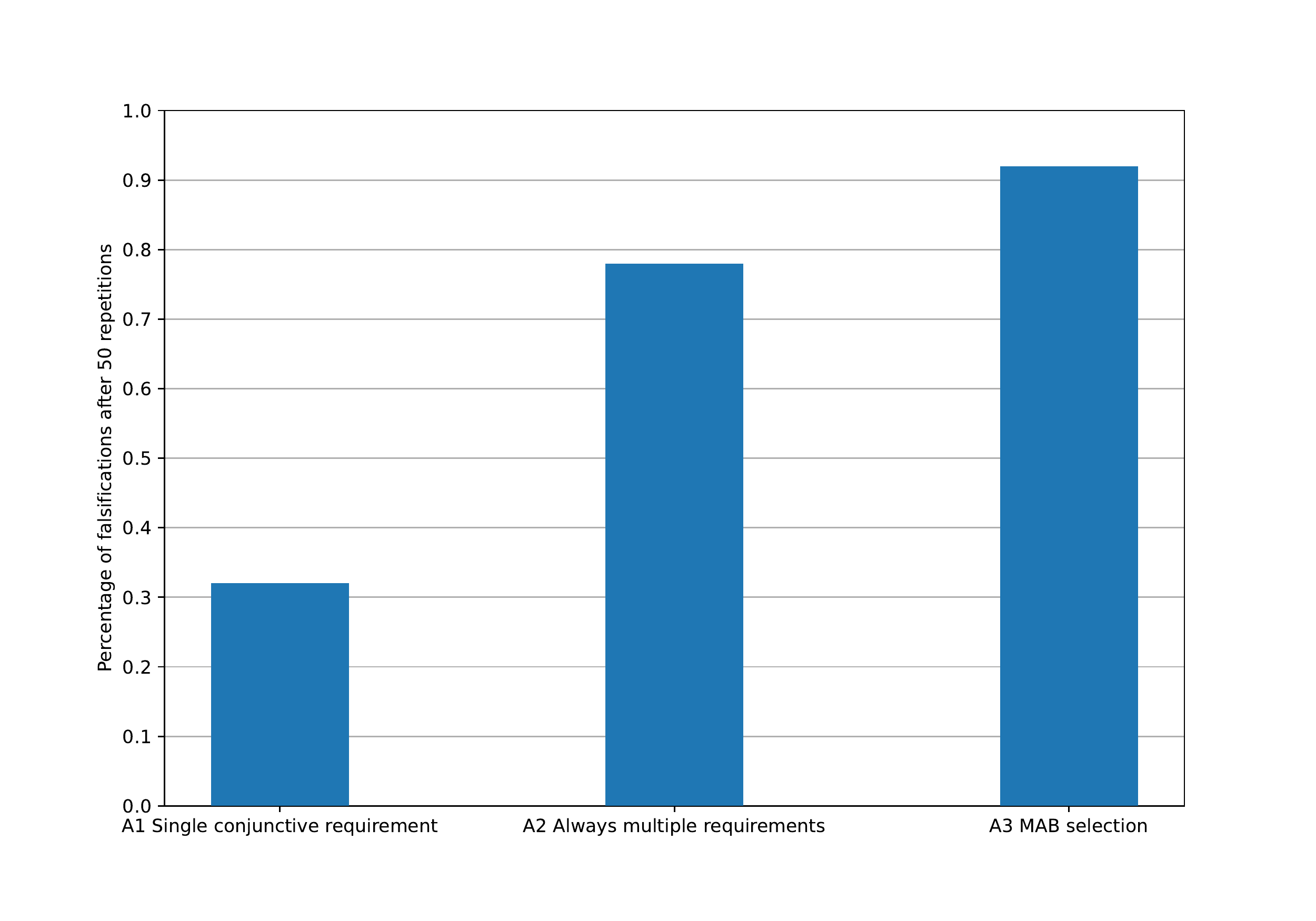}
  \caption{\textmd{Experiment 2: Percentage of successful falsifications.}}\label{fig:mo3d-barp}
\end{figure}

The results written in \autoref{tbl} show that \autoref{alg1} succeeds in falsifying the given requirement only in $16$ of
the $50$ repetitions while Algorithms \ref{alg2} and \ref{alg3} succeed $39$ and $46$ times respectively. These
correspond to $32 \%$, $78 \%$, and $92 \%$ of the repetitions. The success ratios are represented visually in
\autoref{fig:mo3d-barp}. The two proportion Z-test reports that the differences in falsification success rates between
Algorithms \ref{alg2} and \ref{alg3} are statistically significant albeit with an observed $p$-value of $0.05$.

\begin{figure}
  \includegraphics[width=\columnwidth]{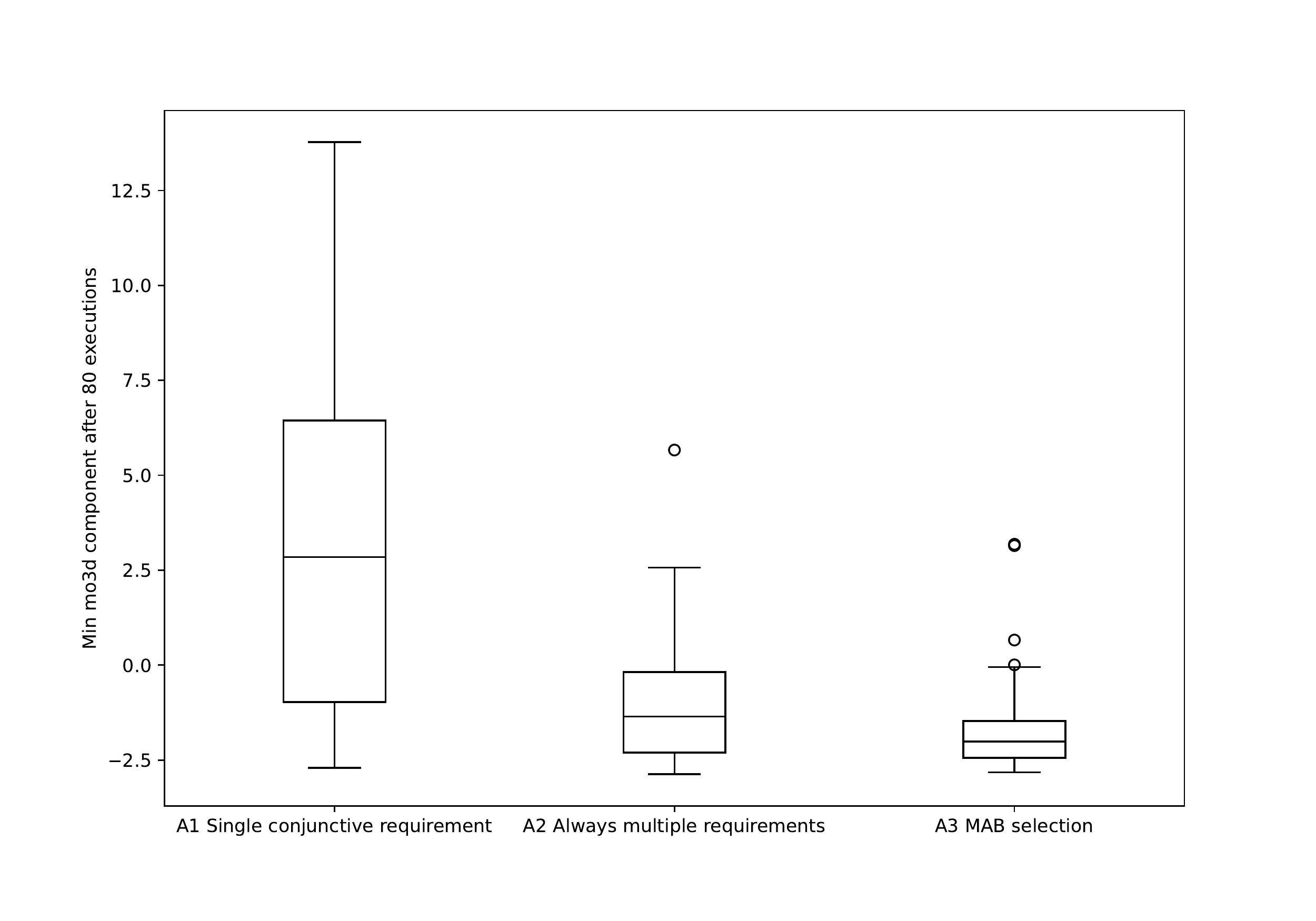}
  \caption{\textmd{Experiment 2: Box plots for minimum $\textup{mo3d}$ component over $50$ experiments.}}\label{fig:mo3d-bp}
\end{figure}

\autoref{fig:mo3d-bp} shows box plots for the minima found by each algorithm after $80$ executions on the SUT. It
clearly shows again that Algorithms \ref{alg2} and \ref{alg3} perform better than \autoref{alg1}. While \autoref{alg1}
produces a median greater than 0, Algorithms \ref{alg2} and \ref{alg3} exhibit a median smaller than $0$. When
comparing Algorithms \ref{alg2} and \ref{alg3}, the first one yields a median of $-1.36$ while the later yields a median
of $-2.01$. The Wilcoxon signed-rank test reports a $p$-value of $0.02$ under the null hypothesis that the median of
differences is $0$.

\begin{figure}
  \includegraphics[width=\columnwidth]{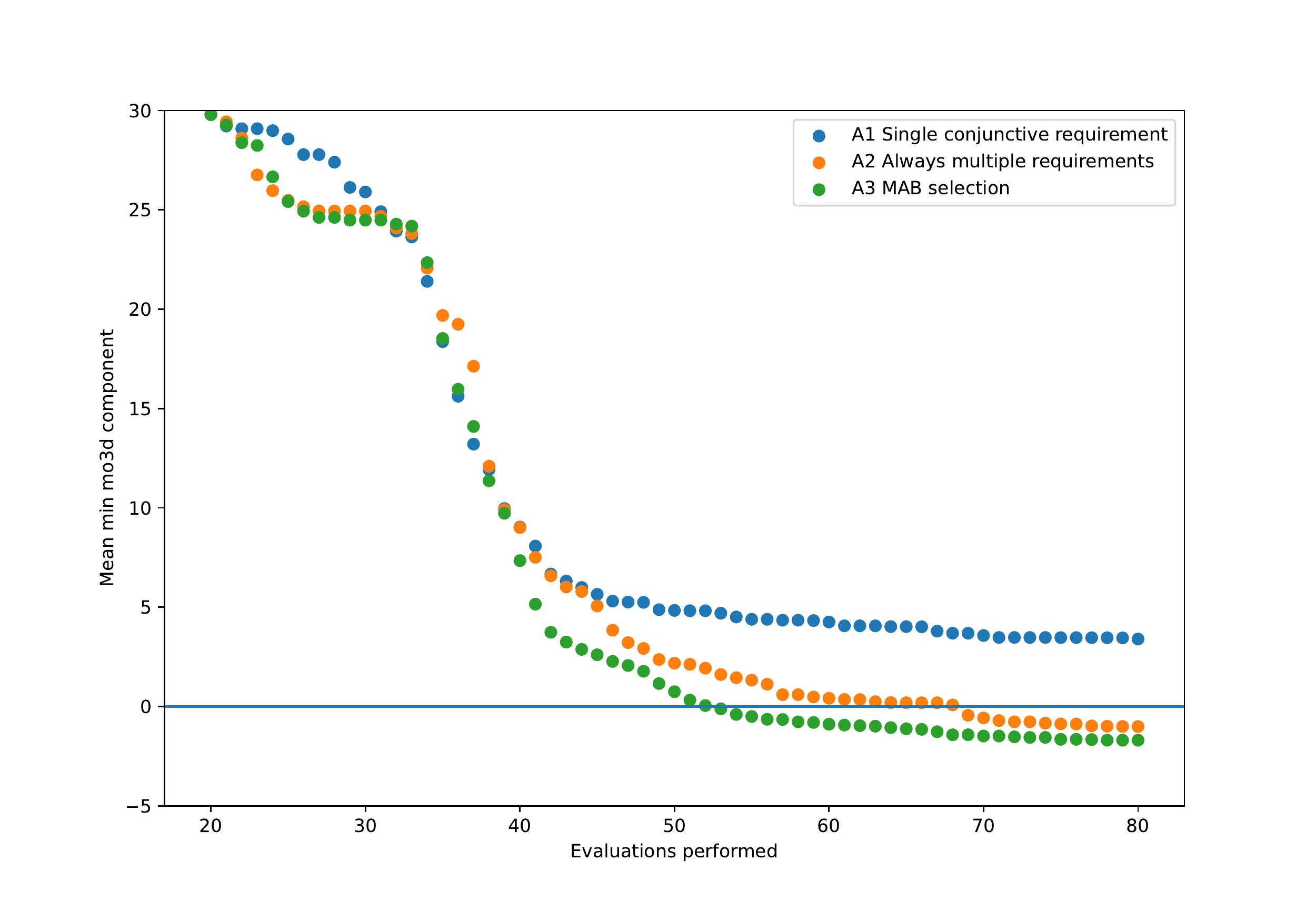}
  \caption{\textmd{Experiment 2: Evolution of minimum $\textup{mo3d}$ component over number of function evaluations, mean of $50$ experiments.}}\label{fig:mo3d-sp}
\end{figure}

Finally, \autoref{fig:mo3d-sp} shows the evolution of the  mean observed minimum over the sequence of function evaluations. We observe how the  mean for the minimum reported by Algorithm 3 becomes smaller than 0 with less evaluations than the other two algorithms.

We have thus empirically confirmed that observing the robustness of each requirement separately can lead to significant
improvement in falsification rate and the number of executions needed for falsification. Moreover the multi-armed
bandit approach of focusing only on the most promising requirement can bring further improvements in falsification in
addition to saving significantly on computational resources.

\section{Conclusions}\label{sec:concl}

In this paper we present a series of algorithms for robustness-based falsification of cyber-physical systems based on the online GAN framework.

The online GAN framework for requirement falsification is a novel concept that apparently has not been proposed before in the research literature. Both our approach and the Bayesian optimization approach of~\cite{DBLP:journals/corr/abs-2005-02979} combine surrogate model creation and test generation in an iterative loop.  One difference is that Bayesian optimization methods use a Gaussian process as a surrogate model while we use a neural network. However we consider that the main difference is that Bayesian optimization methods must query the Gaussian process to determine the next input to evaluate while our approach uses a generative neural network that serves as a model of the space of falsifying inputs. The surrogate model and the generator model are trained together as in a GAN. Our early results indicate that \autoref{alg1}, based on the online GAN framework, exhibits a competitive performance when compared to the results published in~\cite{mathesen2021efficient}.

We also present an extension of our basic algorithm to process conjunctive requirements more efficiently. The approach adopted in \autoref{alg2} is based on the solution provided in~\cite{mathesen2021efficient}. \autoref{alg2} confirms that the idea of unfolding a single conjunctive requirement in multiples requirements can also be used with the online GAN framework. \autoref{alg2} is then extended with the addition of a multi-armed bandit to produce \autoref{alg3}. In our experiment, \autoref{alg3} is more efficient than \autoref{alg2} both in achieving the falsification with less function evaluations and using less computational resources.

We should note that the presented evaluation of the proposed algorithms is based on a  small number of numerical experiments. Also the performance comparisons between the algorithms presented here and the algorithm presented in~\cite{mathesen2021efficient} are only indicative.

As a future work, we acknowledge that a more comprehensive evaluation with a larger set of problems is necessary in order to establish the benefits of the proposed algorithms. Finally, the theoretical underpinnings of the online GAN framework should be studied in more detail in order to understand better its benefits and limitations as an optimization tool.

\section*{Acknowledgments}
This research work has received funding from the ECSEL Joint Undertaking (JU) under grant agreement
No 101007350. The JU receives support from the European Union’s Horizon 2021 research and innovation
programme and Sweden, Austria, Czech Republic, Finland, France, Italy, Spain.
 
\bibliography{iteqs}
\bibliographystyle{plain}

\end{document}